\documentclass[onecolumn,preprint,showpacs,superscriptaddress,nofootinbib]{revtex4-1}

\usepackage[utf8]{inputenc}
\usepackage{mathtools}
\usepackage{graphicx}
\usepackage{caption}
\usepackage{subcaption}
\usepackage{xcolor}
\usepackage{soul}
\usepackage{amsmath,latexsym}

\begin{document}

\title{Rapidly rotating neutron stars in $f(R,T)$ gravity}

\author{F. M. da Silva}
\email{franmdasilva@gmail.com}

\affiliation{Núcleo Cosmo--ufes \& Departamento de Física, Universidade Federal do Espírito Santo, Av. Fernando Ferrari, 540, CEP 29.075-910, Vitória, ES, Brazil}

\author{L. C. N. Santos}
\email{luis.santos@ufsc.br}

\affiliation{Departamento de Física, CCEN--Universidade Federal da Paraíba; C.P. 5008, CEP  58.051-970, João Pessoa, PB, Brazil} 

\author{C. E. Mota}
\email{clesio200915@hotmail.com}

\affiliation{Departamento de Física, CFM--Universidade Federal de Santa Catarina; C.P. 476, CEP 88.040-900, Florianópolis, SC, Brazil} 

\author{T. O. F. da Costa}
\email{tulio.costa@edu.ufes.br}

\affiliation{Núcleo Cosmo--ufes \& Departamento de Física, Universidade Federal do Espírito Santo, Av. Fernando Ferrari, 540, CEP 29.075-910, Vitória, ES, Brazil} 

\author{J. C. Fabris}
\email{julio.fabris@cosmo-ufes.org}

\affiliation{Núcleo Cosmo--ufes \& Departamento de Física, Universidade Federal do Espírito Santo, Av. Fernando Ferrari, 540, CEP 29.075-910, Vitória, ES, Brazil} 

\affiliation{National Research Nuclear University MEPhI, Kashirskoe sh.31, Moscow 115409, Russia}

\begin{abstract}
In this work, we study the influence of $f(R,T)$ gravity on rapidly rotating neutron stars. First we discuss the main aspects of this modified theory of gravity where the gravitational Lagrangian is an arbitrary function of the Ricci scalar $R$ and of the trace of the energy--momentum tensor $T$. Then we present the basic equations for neutron stars including the equations of state used in the present work to describe the hadronic matter. Some physical quantities of interest are calculated such as mass--radius relations, moments of inertia, angular momentum, and compactness. By considering four different rotation regimes, we obtain results that indicate substantial modifications in the physical properties of neutron stars in $f(R,T)$ gravity when compared to those in the context of general relativity. In particular, the mass--radius relation for sequences of stars indicates that $f(R,T)$ gravity increases the mass and the equatorial radius of the neutron stars for stars rotating with an angular velocity smaller than Kepler limit.
\end{abstract}

\maketitle

\section{Introduction}

General relativity (GR) describes the space--time geometric structure based on the content of energy and matter contained within it. On this regard, there are many experimental tests in which GR theory has been confirmed considering  geometries of the space--time associated with different astrophysical systems. More recently, measurements of gravitational waves by the collaboration Virgo and LIGO (Laser Interferometer Gravitational--Wave Observatory) \cite{abbott2016observation,abbott2019gwtc} and the first images of black holes by the project Event Horizon Telescope \cite{akiyama2019first,akiyama2022first} have confirmed predictions of the theory. On the other hand, there are still open problems in the field of cosmology and astrophysics regarding the dark energy and dark matter not yet well understood in the context of GR. In this way, modifications of GR have been developed due to the aforementioned issues. In addition, there are formal aspects associated to the quantization of gravitational field that can be addressed in the framework of modified theories of gravity. Considering the divergences that arise in the context of renormalization at one loop, DeWitt observed that the action for gravity
should be constructed with higher--order curvature terms \cite{witt}. In this way, extensions of GR as $f(R,T)$  gravity \cite{Harko:2011kv} can provide an action for gravity with this requirement since that the theory is formulated with arbitrary functions of curvature terms.

In a cosmological context, the $f(R,T)$ theory was considered in the study of the evolution of scalar cosmological perturbations \cite{alvarenga2013dynamics} and cosmological models with different equations of state (EoS) \cite{jamil2012reconstruction,sharif2012thermodynamics,houndjo2012reconstruction,shabani2014cosmological,moraes2017simplest}.
This modified theory of gravity can be  linked to noncommutative quantum theory by energy nonconservation of the energy--momentum tensor in $f(R,T)$ context \cite{lobato2019energy}. The connection between Rastall gravity and $f(R,T)$ gravity was studied in \cite{shabani2020connection}, where the authors considered that the matter content is a perfect fluid with linear EoS. 

In an astrophysical context, neutron stars (NS) have been used  for the study of the effects of modified theories of gravity due to the very high mass density of these compact objects. In the case of static stars, we can mention some works that consider the influence of different modifications of GR \cite{paper2,paper1,harada1998neutron,orellana2013structure,momeni2015tolman,oliveira2015neutron,hendi2016modified,singh2019einstein,maurya2020charged,mota2019combined,fabris2021stellar}. In particular, the theory of $f(R,T)$ gravity has been studied in several papers that explore effects of this gravity theory on different types of stars \cite{moraes2016stellar,yousaf2016causes,carvalho2017stellar,deb2019study,dos2019conservative,bhatti2019stability,lobato2020neutron,rahaman2020anisotropic}. 

Currently, almost all of the accurate measurements that we have of NS mass and radius come from rotating stars in binary pulsar systems \cite{ozel2016masses}. There are evidences that NS can rotate at an angular velocity of up to $716 $ Hz \cite{hessels2006radio}. Besides, some theoretical models indicate that magnetars can rotate even faster \cite{uso1992millisecond,metzger2011protomagnetar,giacomazzo2013formation,kumar2015physics}. In this way, it is important to explicitly include effects of fast rotation in models of NS. Indeed, rapidly rotating NS have been approached in the context of modified theories of gravity such as scalar--tensor theories of gravity \cite{doneva2013rapidly,doneva2018differentially,astashenok2020rotating}, $f(R)$ gravity  \cite{doneva2015iq,yazadjiev2015rapidly}, dilatonic Einstein--Gauss--Bonnet theory \cite{kleihaus2016rapidly} and in Rastall's gravity \cite{da2021rapidly}.

In this paper, we consider the effects of $f(R,T)$ gravity on rapidly rotating NS. To take into account a star with a rapid rotation, we use the method developed by Komatsu, Eriguchi and Hachisu (KEH)  \cite{komatsu1989rapidly,komatsu1989rapidlyii}. The organization of the paper has the following structure: In Sec. \ref{frt}, we discuss the main aspects of the $f(R,T)$ gravity, and in Sec. \ref{rap} we obtain equations that describe a rapidly spinning star. In Sec. \ref{eos}, we present the EoS used to describe hadronic matter in this paper. The results obtained in our work are in Sec. \ref{results} and the conclusions are in Sec. \ref{conclusions}.    


\section{$f(R,T)$ gravity} \label{frt}

In order to study rapidly rotating NS in $f(R,T)$ gravity, we start by briefly discussing the main aspects of this modified theory of gravity.

In 2011, T. Harko et al. \cite{Harko:2011kv} proposed the theory of gravity $f(R,T)$, where the gravitational Lagrangian is an arbitrary function of the Ricci scalar $R$ and of the trace of the energy--momentum tensor $T$. The dependence on $T$ is justified by the presence of imperfect exotic fluids and quantum effects. To obtain the field equations, we proceed with the generalization $R \rightarrow f(R,T)$ in the Hilbert--Einstein Lagrangian and thus we obtain
\begin{equation}
S=\frac{1}{16 \pi}\int d^{4}xf(R,T)\sqrt{-g}+\int d^{4}x\mathcal{L}_m\sqrt{-g},
\label{action}
\end{equation} 
where $g$ is the determinant of the metric tensor $g_{\mu\nu}$. We will work with geometrized units, that is, $c=G=1$ and we use the metric signature $(-,+,+,+)$. $\mathcal{L}_{m}$ is the Lagrangian density of matter such that we define the energy--momentum tensor of matter as \cite{Landau:1975pou}
\begin{equation}
 T_{\mu\nu} = - \frac{2}{\sqrt{-g}}\frac{\delta(\sqrt{-g}\mathcal{L}_{m})}{\delta g^{\mu\nu}}.   
\end{equation}
Now, assuming that the Lagrangian density $\mathcal{L}_{m}$ depends only on the components $g_{\mu\nu}$ and not on its derivatives, we have
\begin{equation}
   T_{\mu\nu} = g_{\mu\nu}\mathcal{L}_{m} - 2\frac{\partial \mathcal{L}_{m}}{\partial g^{\mu\nu}}.
\end{equation}
By varying the equation (\ref{action}) with respect to $g^{\mu\nu}$, we obtain the field equations of $f(R,T)$ gravity in the metric formalism as
\begin{equation}
\begin{split}
f_R(&R,T)R_{\mu\nu}-\frac{1}{2}f(R,T)g_{\mu\nu}+(g_{\mu\nu} \Box-\nabla_\mu\nabla_\nu)f_R(R,T) \\
&=8 \pi T_{\mu\nu}-f_T(R,T)T_{\mu\nu}-f_T(R,T)\Theta_{\mu\nu},
\end{split}
\label{eq_field}
\end{equation}
where, as originally proposed by the authors, $f_{R}(R,T) \equiv \partial f(R,T)/ \partial R$, $f_{T}(R,T) \equiv \partial f(R,T)/ \partial T$, $\Box \equiv \partial_{\mu}(\sqrt{-g}g^{\mu\nu}\partial_{\nu})/\sqrt{-g}$ and the tensor $\Theta_{\mu\nu}$ is defined as \cite{Harko:2011kv}
\begin{equation}
    \Theta_{\mu\nu} \equiv g^{\alpha\beta}\frac{\delta T_{\alpha\beta}}{\delta g^{\mu\nu}} = -2T_{\mu\nu} + g_{\mu\nu}\mathcal{L}_{m} - 2g^{\alpha\beta}\frac{\partial^{2}\mathcal{L}_{m}}{\partial g^{\mu\nu} \partial g^{\alpha\beta}}.
    \label{eq_tensor}
\end{equation}
Note that when $f(R,T) =f(R)$, from equations (\ref{eq_field}) we obtain the field equations in the context of $f(R)$ gravity. At this stage, taking the covariant divergence of (\ref{eq_field}), with the use of mathematical identity \cite{Koivisto:2005yk}
\begin{equation}
    \nabla^{\mu}\left[f_{R}(R,T)R_{\mu\nu} -\frac{1}{2}f(R,T)g_{\mu\nu}+(g_{\mu\nu} \Box-\nabla_\mu\nabla_\nu)f_R(R,T) \right] \equiv 0,
\end{equation}
we find the following modified equation for the four--divergence of the energy--momentum tensor $T_{\mu\nu}$ \cite{BarrientosO:2014mys}
\begin{equation}
\begin{split}
\nabla^{\mu}T_{\mu\nu}=&\frac{f_T(R,T)}{8 \pi  -f_T(R,T)}\left[(T_{\mu\nu}+\Theta_{\mu\nu})\nabla^{\mu}\ln f_T(R,T) \right. \\
& \left. +\nabla^{\mu}\Theta_{\mu\nu}-\frac{1}{2}g_{\mu\nu}\nabla^{\mu}T\right].
\label{eq_conserv}
\end{split}
\end{equation}

\section{Rapidly rotating star in $f(R,T)$ gravity} \label{rap}

Now, we will obtain the equations that describe a stationary rapidly spinning star in $f(R,T)$ gravity. We start by writing the line element that describes the space--time geometry of a stationary axisymmetric rotating star, which can be written in terms of spherical coordinates $\left(t,r,\theta,\phi\right)$ as follows:
\begin{equation}
ds^{2}=-e^{\gamma + \rho }dt^{2}+e^{2\alpha }\left( dr^{2}+r^{2}d\theta ^{2}\right)
+e^{\gamma - \rho}r^{2}\sin ^{2}\theta \left( d\phi -\omega dt\right) ^{2},
\label{rs1}
\end{equation}
where $\alpha$, $\gamma$, $\rho$ and $\omega$ are the metric functions which depend only on $r$ and $\theta$, and the function $\omega$ is associated to the Lense--Thirring effect \cite{pfister2007history,ciufolini2004confirmation}.  

We assume that the matter of which the star is made is a perfect fluid. 
In this case, there is no unique definition of the matter Lagrangian density. However, in the present study we assume that $\mathcal{L}_{m} = p$, with $p$ being the pressure of the fluid \cite{schutz1970perfect}. In this case the tensor $\Theta_{\mu\nu}$ is given by
\begin{equation}
\Theta_{\mu\nu}=-2T_{\mu\nu}+pg_{\mu\nu}.
\end{equation}
And for the energy--momentum tensor, we use the general definition \cite{Fisher:2019ekh} given by the following expression
\begin{equation}
    T_{\mu\nu}=pg_{\mu\nu}+(p + \varepsilon)U_{\mu}U_{\nu},
\end{equation}
where $\varepsilon$ is the energy density and $U_{\mu}$ represents the 4--velocity of the fluid, which obeys the conditions: $U_{\mu}U^{\mu}=-1$ and $U^{\mu}\nabla_{\nu}U_{\mu}=0$. In the case of a stationary rotating star the expression for $U^{\mu}$ is given by \cite{komatsu1989rapidly}:
\begin{equation}
U^{\mu}=\frac{dx^{\mu}}{d\tau }=\frac{e^{-\frac{\gamma+\rho}{2} }}{\sqrt{1-v^{2}}}\left(
1,0,0,\Omega \right), \label{rs3}
\end{equation}
where $\Omega$ represents the angular velocity of an element of mass of the star with respect to a static observer at infinity, and $v$ is the 3--velocity in the ZAMO (zero momentum angular observer), which is the referential frame of an observer locally without rotation, given by
\begin{equation}
v=\left( \Omega -\omega \right) r\sin \theta e^{-\rho}.   \label{rs4}
\end{equation}

For the functional form of the $f(R,T)$ function, we use one that was originally suggested by T.Harko et al. in \cite{Harko:2011kv}, {\it i.e}, $f(R,T)=R+2\lambda' T$, with $\lambda'$ a constant. The substitution of $f(R,T)=R+2\lambda' T$ in equations (\ref{eq_field}) and (\ref{eq_conserv}) give us
\begin{equation}
G_{\mu\nu}=(8 \pi +2\lambda') T_{\mu\nu}+\lambda' g_{\mu\nu}(T-2p),
\label{eq_likeEinstein}
\end{equation}
and
\begin{equation}
\nabla^{\mu}T_{\mu\nu}=\frac{2\lambda'}{8 \pi +2\lambda'}\left[\nabla^{\mu}(pg_{\mu\nu})-\frac{1}{2}g_{\mu\nu}\nabla^{\mu}T\right], \label{tov_like1}
\end{equation} 
where $G_{\mu\nu}$ is the usual Einstein tensor, so that the left side of equation (\ref{eq_likeEinstein}) is equal to the left side of Einstein's equations of GR, and in the right side we have additional terms proportional to the parameter $\lambda'$ of the $f(R,T)$ theory.

Now, we can define an effective energy--momentum tensor given by
\begin{equation}
    \tau_{\mu\nu}=T_{\mu\nu} + \frac{\lambda'}{8 \pi +2 \lambda'}g_{\mu\nu}(T-2p),
\end{equation}
and using the definition above we rewrite the field equations (\ref{eq_likeEinstein}) in a compact form as follows
\begin{equation}
G_{\mu\nu}=8 \pi (1 + 2\lambda) \tau_{\mu\nu},
\label{eq_field_eff}
\end{equation}
where $\lambda= \lambda' / (8 \pi )$. With such formalism, the application of the Bianchi identities in the above equation yields
\begin{equation}
\nabla^{\mu}\tau_{\mu\nu}=0. \label{tov_like2}
\end{equation} 
It is interesting to note that for a perfect fluid the effective energy--momentum tensor  $\tau_{\mu\nu}$  can be written as
\begin{equation}
\tau_{\mu\nu}=p^{eff} g_{\mu\nu} + (p^{eff}+\varepsilon^{eff})U_{\mu}U_{\nu}, \label{tminieff}
\end{equation}
where $\varepsilon^{eff}$ is the effective energy density and $p^{eff}$ is the effective pressure as follows
\begin{equation} \label{eff}
\begin{split}
& \varepsilon^{eff} = \frac{\varepsilon(1+3\lambda)-p\lambda}{1+2\lambda}, \\
& p^{eff} = \frac{p(1+3\lambda)-\varepsilon\lambda}{1+2\lambda}.
\end{split}
\end{equation}
We can verify that equations (\ref{tov_like1}) and (\ref{tov_like2}) are equivalent.

Applying the metric (\ref{rs1}) and the effective energy--momentum tensor (\ref{tminieff}) into (\ref{eq_field_eff}) give us the following field equations:
\begin{equation}
\nabla^{2} \left( \rho e^{\gamma /2}\right) =S_{\rho }\left( r,\mu \right),
\label{rs6}
\end{equation}
\begin{equation}
\left( \nabla^{2} +\frac{1}{r}\frac{\partial }{\partial r}-\frac{1}{r^{2}}\mu
\frac{\partial }{\partial \mu }\right) \gamma e^{\gamma /2}=S_{\gamma
}\left( r,\mu \right),  \label{rs7}
\end{equation}
\begin{equation}
\left( \nabla^{2} +\frac{2}{r}\frac{\partial }{\partial r}-\frac{2}{r^{2}}\mu
\frac{\partial }{\partial \mu }\right) \omega e^{\left( \gamma -2\rho
\right) /2}=S_{\omega }\left( r,\mu \right)  \label{rs8}
\end{equation}
and
\begin{equation}
\begin{split}
\alpha _{,\mu }=& -\nu _{,\mu }-\left\{ \left( 1-\mu
^{2}\right) \left( 1+rB^{-1}B_{,r}\right) ^{2}+\left[ \mu -\left(
1-\mu ^{2}\right) B^{-1}B_{,\mu }\right] ^{2}\right\} ^{-1} \\
& \left[ \frac{1}{2}B^{-1}\left\{ r^{2}B_{,rr}-\left[ \left(
1-\mu ^{2}\right) B_{,\mu }\right] _{,\mu }-2\mu
B_{,\mu }\right\} \left[ -\mu +\left( 1-\mu ^{2}\right)
B^{-1}B_{,\mu }\right] \right.  \\
& +rB^{-1}B_{,r}\left[ \frac{1}{2}\mu +\mu rB^{-1}B_{,r}+%
\frac{1}{2}\left( 1-\mu ^{2}\right) B^{-1}B_{,\mu }\right]  \\
& +\frac{3}{2}B^{-1}B_{,\mu }\left[ -\mu ^{2}+\mu \left( 1-\mu
^{2}\right) B^{-1}B_{,\mu }\right] -\left( 1-\mu ^{2}\right)
rB^{-1}B_{,\mu r} \\
&  (1+rB^{-1}B_{,r})-\mu r^{2}\left( \nu _{,r}\right)
^{2}-2\left( 1-\mu ^{2}\right) r\nu _{,\mu }\nu _{,r}+\mu
\left( 1-\mu ^{2}\right) \left( \nu _{,\mu }\right) ^{2} \\
& -2\left( 1-\mu ^{2}\right) r^{2}B^{-1}B_{,r}\nu _{,\mu
}\nu _{,r}+\left( 1-\mu ^{2}\right) B^{-1}B_{,\mu }\left[
r^{2}\left( \nu _{,r}\right) ^{2}\right. \\
&  \left.-\left( 1-\mu ^{2}\right) \left(
\nu _{,\mu }\right) ^{2}\right] +\left( 1-\mu ^{2}\right) B^{2}e^{-4\nu }\left\{ \frac{1}{4}\mu
r^{4}\left( \omega _{,r}\right) ^{2}+\frac{1}{2}\left( 1-\mu
^{2}\right) r^{3}\omega _{,\mu }\omega _{,r}\right.  \\
& -\frac{1}{4}\mu \left( 1-\mu ^{2}\right) r^{2}\left( \omega _{,\mu }\right) ^{2}+\frac{1}{2}\left( 1-\mu ^{2}\right)
r^{4}B^{-1}B_{,r}\omega _{,\mu }\omega _{,r} \\
& \left. \left. -\frac{1}{4}\left( 1-\mu ^{2}\right) r^{2}B^{-1}B_{,\mu }\left[ r^{2}\left( \omega _{,r}\right) ^{2}-\left( 1-\mu
^{2}\right) \left( \omega_{,\mu}\right) ^{2}\right] \right\} %
\right],
\end{split}
\label{rs21}
\end{equation}
where $\mu$ is defined as
\begin{equation}
    \mu=\cos\theta, \label{rs9}
\end{equation}
the function $B$ is as follows
\begin{equation}
    B=e^{\gamma } \label{rs22}
\end{equation}
and the operator $\nabla^2$ is given by
\begin{equation}
\nabla^{2} =\frac{\partial ^{2}}{\partial r^{2}}+\frac{2}{r}\frac{\partial }{%
\partial r}+\frac{1}{r^{2}}\frac{\partial}{\partial \mu}\left((1-\mu^{2})\frac{\partial}{\partial\mu}\right)+\frac{1}{r^{2}(1-\mu^{2})}\frac{\partial ^{2}}{\partial \phi ^{2}}. \label{rs10}
\end{equation}
The ``source" terms that appear in equations $(\ref{rs6})$, $(\ref{rs7})$ and $(\ref{rs8})$ are given by
\begin{equation}
\begin{split}
S_{\rho }(r,\mu )=& \left[ e^{\gamma /2} k e^{2\alpha }\left( \varepsilon
+p\right) \frac{1+v^{2}}{1-v^{2}}\right.  \\
& +r^{2}(1-\mu ^{2})e^{-2\rho }\left[ \omega _{,r}^{2}+\frac{1}{%
r^{2}}\left( 1-\mu ^{2}\right) \omega _{,\mu }^{2}\right] +\frac{1}{%
r}\gamma _{,r}-\frac{1}{r^{2}}\mu \gamma _{,\mu } \\
& \left. +\frac{\rho }{2}\left\{ 2 k e^{2\alpha }\frac{p(1+3\lambda)-\varepsilon\lambda}{1+2\lambda}-\gamma _{,r}\left( \frac{1}{2}\gamma _{,r}+\frac{1}{r}\right) -\frac{1}{r^{2}%
}\gamma _{,\mu }\left[ \frac{1}{2}\gamma _{,\mu }\left(
1-\mu ^{2}\right) -\mu \right] \right\} \right],
\end{split}
\end{equation}
\begin{equation}
S_{\gamma }\left( r,\mu \right) =e^{\gamma /2}\left\{ 2 k e^{2\alpha }\frac{p(1+3\lambda)-\varepsilon\lambda}{1+2\lambda}+%
\frac{\gamma }{2}\left[ 2k e^{2\alpha }\frac{p(1+3\lambda)-\varepsilon\lambda}{1+2\lambda}-\frac{1}{2}\gamma _{,r}^{2}-\frac{1}{2r^{2}}\left( 1-\mu ^{2}\right) \frac{1}{2}\gamma
_{,\mu }^{2}\right] \right\}
\end{equation}
and
\begin{equation}
\begin{split}
S_{\omega }\left( r,\mu \right) =& e^{\left( \gamma -2\rho \right) /2}\left[
-2 k e^{2\alpha }\frac{\left( \Omega -\omega \right) \left( \varepsilon
+p\right) }{1-v^{2}}\right.  \\
& +\omega \left\{ - k e^{2\alpha }\left[\frac{ \left( 1+v^{2}\right)
\left(\varepsilon +p\right)}{1-v^{2}}-\frac{p(1+3\lambda)-\varepsilon\lambda}{1+2\lambda}\right] -\frac{1}{r}\left( 2\rho _{,r}+\frac{1}{2}\gamma _{,r}\right) \right.  \\
& +\frac{1}{r^{2}}\mu \left( 2\rho _{,\mu }+\frac{1}{2}\gamma
_{,\mu }\right) +\frac{1}{4}\left( 4\rho _{,r}^{2}-\gamma
_{,r}^{2}\right) +\frac{1}{4r^{2}}\left( 1-\mu ^{2}\right) \left(
4\rho _{,\mu }^{2}-\gamma _{,\mu }^{2}\right)  \\
& \left. \left. -r^{2}\left( 1-\mu ^{2}\right) e^{-2\rho }\left[ \omega
_{,r}^{2}+\frac{1}{r^{2}}\left( 1-\mu ^{2}\right) \omega _{,\mu }^{2}\right] \right\} \right].
\end{split}
\end{equation}
respectively, with $k=8 \pi (1+2\lambda)$.

In this work we use the scheme developed by KEH \cite{komatsu1989rapidly,komatsu1989rapidlyii}, which is an iterative numerical method that uses suitable Green functions to obtain the following integral representation of the equations $(\ref{rs6})$, $(\ref{rs7})$ and $(\ref{rs8})$:
\begin{equation}
\rho =-e^{-\gamma /2}\sum_{n=0}^{\infty }P_{2n}\left( \mu \right)
\int_{0}^{\infty }r^{\prime 2}f_{2n}^{2}\left( r,r^{\prime
}\right) \int_{0}^{1}P_{2n}\left( \mu ^{\prime }\right)
S_{\rho }\left( r^{\prime },\mu ^{\prime }\right)d\mu^{\prime } dr^{\prime},  \label{rs16}
\end{equation}
\begin{equation}
\gamma  =-\frac{2e^{-\gamma /2}}{\pi r\sin \theta }\sum_{n=1}^{\infty }%
\frac{\sin{\left( 2n-1\right) \theta} }{2n-1}\int_{0}^{\infty }r^{\prime 2}f_{2n-1}^{1}\left( r,r^{\prime }\right) \int_{0}^{1}\sin{\left( 2n-1\right) \theta^{\prime}}S_{\gamma }\left( r^{\prime },\mu ^{\prime }\right)d\mu^{\prime } dr^{\prime},
\label{rs17}
\end{equation}
and
\begin{equation}
\begin{split}
\omega & =-\frac{e^{\left( 2\rho -\gamma \right) /2}}{r\sin \theta }%
\sum_{n=1}^{\infty }\frac{P_{2n-1}^{1}\left( \mu \right) }{2n\left(
2n-1\right) }\int_{0}^{\infty }r^{\prime 3}f_{2n-1}^{2}\left(
r,r^{\prime }\right) \int_{0}^{1}P_{2n-1}^{1}\left( \mu ^{\prime }\right)
S_{\omega }\left( r^{\prime },\mu ^{\prime }\right) d\mu^{\prime } dr^{\prime},
\end{split}
\label{rs18}
\end{equation}
where $P_{n}$ are the Legendre polynomials, $P^{m}_{n}$ are the Associated Legendre functions and the functions $f_{n}^{1}\left( r,r^{\prime }\right)$ and $f_{n}^{2}\left( r,r^{\prime }\right)$ are given by
\begin{equation}
f_{n}^{1}\left( r,r^{\prime }\right) =\left\{
\begin{tabular}{l}
${\left( r^{\prime }/r\right) ^{n}\text{, if }r^{\prime }\leq r,}$ \\
${\left( r/r^{\prime }\right) ^{n}\text{, if }r^{\prime }>r,}$%
\end{tabular}
\right.  \label{rs19}
\end{equation}
\begin{equation}
f_{n}^{2}\left( r,r^{\prime }\right) =\left\{
\begin{tabular}{l}
${r^{\prime n}/r^{n+1}\text{, if }r^{\prime }\leq r,}$ \\
${r^{n}/r^{\prime n+1}\text{, if }r^{\prime }>r}$.
\end{tabular}
\right.  \label{rs20}
\end{equation}
The equation (\ref{rs21}) for the function $\alpha$ is the only one that is not transformed to a integral equation, it can be integrated using the following condition,
\begin{equation}
    \alpha(r,1)= \frac{\gamma(r,1) -\rho(r,1)}{2}. \label{rs23}
\end{equation}
A convenient feature of the KEH method is that the requirement that the metric functions $\rho$, $\gamma$ and $\omega$ be asymptotically flat at infinity is promptly satisfied on the condition that the ``source" terms $S_{\rho}$, $S_{\gamma}$ and $S_{\omega}$ are well behaved. As for the function $\alpha$, its flatness condition is automatically satisfied because the other functions already satisfy their respective conditions.  

Computing equation (\ref{tov_like1}), or equivalently equation (\ref{tov_like2}), for divergence of the energy--momentum tensor lead us to the following expression
\begin{equation}
\frac{1}{(1+2\lambda)(p + \varepsilon)}\left[\left(1+3\lambda\right)\nabla p-\lambda\nabla \varepsilon\right]=\nabla \ln U^t,
\end{equation}
that can be integrated to obtain 
\begin{equation} 
h^{eff} = \log(c) -\frac{\gamma+\rho}{2} - \frac{1}{2} \log(1-v^2), \label{heff}
\end{equation}
where $c$ is a constant of integration and $h^{eff}$ can be referred as effective specific enthalpy and is given by the equation bellow
\begin{equation}
\frac{dh^{eff}}{dp^{eff}} = \frac{1}{p^{eff}+\varepsilon^{eff}} , \label{heff2}
\end{equation}
where $p^{eff}$ and $\varepsilon^{eff}$ are defined in equation (\ref{eff}). To be able to solve equation (\ref{heff}), (\ref{heff2}) we need an EoS, and so in the next section we present the EoS we use in this work.

\section{Equations of state \label{eos}}

In this section, we outline the EoS used in the present work to describe hadronic matter. These EoS will be the input to the stellar structure equations that govern rapidly rotating NS in $f(R,T)$ gravity. For a didactic and more extensive explanation on EoS, we refer the reader to ref. \cite{Menezes2021} and references therein.

\subsection{The Nonlinear Walecka Model}

The first relativistic model used here to describe the hadronic matter is a rather generalized version of the quantum hadrodynamics (QHD) \cite{livrowalecka,walecka,bogutabodmer}, which is based on a relativistic mean--field theory and describes the baryon interaction through the exchange of scalar and vector mesons, known as nonlinear Walecka model (NLWM).  

Although the first version of the model \cite{Walecka-74} reproduced well--established properties of infinite nuclear matter, such as, the binding energy and the nuclear saturation density ($\rho_{0} \sim $ 0.15 fm$^{-3}$), other important properties such as incompressibility and the effective mass of nucleons are not obtained with satisfactory values. This problem was circumvented with the introduction of self--interacting terms, cubic and quartic, in the scalar field by Boguta and Bodmer \cite{bogutabodmer}. Likewise, to deal with asymmetric systems with respect to the numbers of protons and neutrons, the vector--isovector meson $\rho$ (not to confound with the notation for the density) was introduced. And, to adjust other properties such as symmetry energy and the fact that protons and neutrons have slightly different masses, other mesons and interactions were included, leading to extensive generalizations and parameterization of this model \cite{Dutra2014}. Therefore, the more general Lagrangian density of the NLWM model is given by \cite{Menezes2021,Agrawal2010,Dutra2014}:

\begin{eqnarray}
\mathcal{L} = \mathcal{L}_{\rm nm} + \mathcal{L}_\sigma +
\mathcal{L}_\omega
+ \mathcal{L}_\rho + \mathcal{L}_{\delta} + \mathcal{L}_{\sigma\omega\rho},
\label{dl}
\end{eqnarray}
where 
\begin{widetext}

\begin{eqnarray}
\mathcal{L}_{\rm nm} & =& \overline{\psi}(i\gamma^\mu\partial_\mu - M)\psi 
+ g_\sigma\sigma\overline{\psi}\psi 
- g_\omega\overline{\psi}\gamma^\mu\omega_\mu\psi 
- \frac{g_\rho}{2}\overline{\psi}\gamma^\mu\vec{\rho}_\mu\cdot \vec{\tau}\psi
+ g_\delta\overline{\psi}\vec{\delta}\cdot \vec{\tau}\psi, 
\label{lag1} \\
\mathcal{L}_\sigma &=& \frac{1}{2}(\partial^\mu \sigma \partial_\mu \sigma 
- m^2_\sigma\sigma^2) - \frac{A}{3}\sigma^3 - \frac{B}{4}\sigma^4,
\\
\mathcal{L}_\omega &=& -\frac{1}{4}F^{\mu\nu}F_{\mu\nu} 
+ \frac{1}{2}m^2_\omega\omega_\mu\omega^\mu 
+ \frac{C}{4}(g_\omega^2\omega_\mu\omega^\mu)^2,
\\
\mathcal{L}_\rho &=& -\frac{1}{4}\vec{B}^{\mu\nu}\vec{B}_{\mu\nu} 
+ \frac{1}{2}m^2_\rho\vec{\rho}_\mu \cdot \vec{\rho}^\mu,
\\
\mathcal{L}_\delta &=& \frac{1}{2}(\partial^\mu\vec{\delta}\partial_\mu\vec{\delta}
- m^2_\delta\vec{\delta}^2),
\\
\mathcal{L}_{\sigma\omega\rho} &=& 
g_\sigma g_\omega^2\sigma\omega_\mu\omega^\mu
\left(\alpha_1+\frac{1}{2}{\alpha_1}'g_\sigma\sigma\right)
+ g_\sigma g_\rho^2\sigma\vec{\rho}_\mu \cdot \vec{\rho}^\mu
\left(\alpha_2+\frac{1}{2}{\alpha_2}'g_\sigma\sigma\right) 
\nonumber \\
&+& \frac{1}{2}{\alpha_3}'g_\omega^2 g_\rho^2\omega_\mu\omega^\mu
\vec{\rho}_\mu\cdot \vec{\rho}^\mu.
\label{lomegarho}
\end{eqnarray}
\end{widetext}
In this Lagrangian density: $\mathcal{L}_{\rm nm}$ represents the kinetic part of the nucleons plus the terms standing for the interaction between them and mesons $\sigma$, $\delta$, $\omega$, and $\rho$; the terms $\mathcal{L}_j$ represents the free and self--interacting terms of the meson $j$, where $j=\sigma,\delta,\omega,$ and $\rho$. The term $\mathcal{L}_{\sigma\omega\rho}$, accounts for crossing interactions between the meson fields. The antisymmetric field tensors $F_{\mu\nu}$ and $\vec{B}_{\mu\nu}$ are given by 
$F_{\mu\nu}=\partial_\nu\omega_\mu-\partial_\mu\omega_\nu$
and $\vec{B}_{\mu\nu}=\partial_\nu\vec{\rho}_\mu-\partial_\mu\vec{\rho}_\nu
- g_\rho (\vec{\rho}_\mu \times \vec{\rho}_\nu)$. Finally, $M$ and $m_j$ are respectively the nucleon mass and the meson masses \footnote{Observe that the notation used here is valid within Section \ref{eos} only.}.

In the relativistic mean field (RMF) approximation, the meson fields are treated as classical fields, and the equations of motion are obtained using the Euler--Lagrange equations assuming rotational and translational invariance. Therefore, the RMF consists of the application of substitutions
\begin{eqnarray}
\sigma\rightarrow \left<\sigma\right>\equiv\sigma_0, \nonumber \\ 
\omega_\mu\rightarrow \left<\omega_0\right>\equiv\omega_0, \nonumber \\
\vec{\rho}_\mu\rightarrow \left<\vec{\rho}_0\right>\equiv \bar{\rho}_{0(3)}, \nonumber \\ 
\vec{\delta}\rightarrow\,\,<\vec{\delta}>\equiv\delta_{(3)} \nonumber.
\label{meanfield}
\end{eqnarray}
and the equations of motion:

\begin{widetext}
\begin{eqnarray}
&m^2_\sigma\sigma_0 = g_\sigma\rho_s - A\sigma_0^2 - B\sigma_0^3 
+g_\sigma g_\omega^2\omega_0^2(\alpha_1+{\alpha_1}'g_\sigma\sigma)
+g_\sigma g_\rho^2\bar{\rho}_{0(3)}^2(\alpha_2+{\alpha_2}'g_\sigma\sigma)\,\mbox{,}\quad 
\label{sigmaacm}\\
&m_\omega^2\omega_0 = g_\omega\rho - Cg_\omega(g_\omega \omega_0)^3 
- g_\sigma g_\omega^2\sigma_0\omega_0(2\alpha_1+{\alpha_1}'g_\sigma\sigma_0)
- {\alpha_3}'g_\omega^2 g_\rho^2\bar{\rho}_{0(3)}^2\omega_0, 
\label{omegaacm}\\
&m_\rho^2\bar{\rho}_{0(3)} = \frac{g_\rho}{2}\rho_3 
-g_\sigma g_\rho^2\sigma_0\bar{\rho}_{0(3)}(2\alpha_2+{\alpha_2}'g_\sigma\sigma_0)
-{\alpha_3}'g_\omega^2 g_\rho^2\bar{\rho}_{0(3)}\omega_0^2, 
\label{rhoacm} \\
&m_\delta^2\delta_{(3)} = g_\delta\rho_{s3}, 
\label{deltaacm}\\
&\left[i\gamma^\mu \partial_\mu -\gamma^0 V_\tau  - (M+S_\tau)\right] \psi = 0,
\end{eqnarray}
\end{widetext}
where 
\begin{equation}
  \rho_s =\left<\overline{\psi}\psi\right>={\rho_s}_p+{\rho_s}_n,
\end{equation}

\begin{equation}
\rho_{s3}=\left<\overline{\psi}{\tau}_3\psi\right>={\rho_s}_p-{\rho_s}_n,
\label{rhos_tot}
\end{equation}

\begin{equation}
  \rho =\left<\overline{\psi}\gamma^0\psi\right> = \rho_p + \rho_n,
 \end{equation}
 
\begin{equation} 
\rho_3=\left<\overline{\psi}\gamma^0{\tau}_3\psi\right> = \rho_p - \rho_n=(2y_p-1)\rho,
\label{rho_tot}
\end{equation}
with
\begin{equation}
{\rho_s}_{p,n} = \frac{\gamma M^*_{p,n}}{2\pi^2}\int_0^{{k_F}_{p,n}}
\frac{k^2dk}{\sqrt{k^2+M^{*2}_{p,n}}} ,
\label{rhospn}
\end{equation}

\begin{equation}
\rho_{p,n} = \frac{\gamma}{2\pi^2}\int_0^{{k_F}_{p,n}}k^2dk =
\frac{\gamma}{6\pi^2}{k_F^3}_{p,n},
\label{rhopn}
\end{equation}

\begin{equation}
V_{\tau} =g_\omega\omega_0 +
\frac{g_\rho}{2}\bar{\rho}_{0(3)}\tau_3 ,\qquad
S_{\tau} =-g_\sigma\sigma_0 -g_\delta\delta_{(3)}\tau_3.
\end{equation}
where $\gamma$ is the spin degeneracy and ${k_F}_{p,n}$ is the Fermi momentum. The indices $p,n$ correspond to protons and neutrons respectively.

The proton and neutron effective masses are given by:
\begin{equation}
    M^{*}_{p} = M - g_{\sigma}\sigma_{0}-g_{\delta}\delta_{(3)}, \quad M^{*}_{n} = M - g_{\sigma}\sigma_{0}+g_{\delta}\delta_{(3)}.
\end{equation}
Finally, after some analytical calculations it is possible to obtain the energy density and pressure in the NLWM model. These quantities are given as follows \cite{Dutra2014,Menezes2021}: 
\begin{widetext}
\begin{eqnarray}
\mathcal{E} &=& \frac{1}{2}m^2_\sigma\sigma_0^2 
+ \frac{A}{3}\sigma_0^3 + \frac{B}{4}\sigma_0^4 - \frac{1}{2}m^2_\omega\omega_0^2 
- \frac{C}{4}(g_\omega^2\omega_0^2)^2 - \frac{1}{2}m^2_\rho\bar{\rho}_{0(3)}^2
+g_\omega\omega_0\rho+\frac{g_\rho}{2}\bar{\rho}_{0(3)}\rho_3
\nonumber \\
&+& \frac{1}{2}m^2_\delta\delta^2_{(3)} - g_\sigma g_\omega^2\sigma\omega_0^2
\left(\alpha_1+\frac{1}{2}{\alpha_1}'g_\sigma\sigma_0\right) 
- g_\sigma g_\rho^2\sigma\bar{\rho}_{0(3)}^2 
\left(\alpha_2+\frac{1}{2}{\alpha_2}' g_\sigma\sigma_0\right) \nonumber \\
&-& \frac{1}{2}{\alpha_3}'g_\omega^2 g_\rho^2\omega_0^2\bar{\rho}_{0(3)}^2
+ \mathcal{E}_{\mbox{\tiny kin}}^p + \mathcal{E}_{\mbox{\tiny kin}}^n,
\label{denerg}
\end{eqnarray}
\end{widetext}
with
\begin{eqnarray}
\mathcal{E}_{\mbox{\tiny kin}}^{p,n}&=&\frac{\gamma}{2\pi^2}\int_0^{{k_F}_{p,n}}k^2
(k^2+M^{*2}_{p,n})^{1/2}dk,
\label{decinnlw}
\end{eqnarray}
and pressure:
\begin{widetext}
\begin{eqnarray}
P &=& - \frac{1}{2}m^2_\sigma\sigma_0^2 - \frac{A}{3}\sigma_0^3 -
\frac{B}{4}\sigma_0^4 + \frac{1}{2}m^2_\omega\omega_0^2 
+ \frac{C}{4}(g_\omega^2\omega_0^2)^2 + \frac{1}{2}m^2_\rho\bar{\rho}_{0(3)}^2
+ \frac{1}{2}{\alpha_3}'g_\omega^2 g_\rho^2\omega_0^2\bar{\rho}_{0(3)}^2
\nonumber \\
&-&\frac{1}{2}m^2_\delta\delta^2_{(3)} + g_\sigma g_\omega^2\sigma_0\omega_0^2
\left(\alpha_1+\frac{1}{2}{\alpha_1}'g_\sigma\sigma_0\right) 
+ g_\sigma g_\rho^2\sigma\bar{\rho}_{0(3)}^2 
\left(\alpha_2+\frac{1}{2}{\alpha_2}' g_\sigma\sigma\right) \nonumber \\
&+& P_{\mbox{\tiny kin}}^p + P_{\mbox{\tiny kin}}^n,\qquad
\label{pressure}
\end{eqnarray}
\end{widetext}
with
\begin{equation}
P_{\mbox{\tiny kin}}^{p,n} = 
\frac{\gamma}{6\pi^2}\int_0^{{k_F}_{p,n}}\frac{k^4dk}{(k^2+M^{*2}_{p,n})^{1/2}}.
\end{equation}

When stellar matter is considered, charge neutrality and chemical $\beta$ equilibrium equations have to be imposed. To implement them, leptons (generally electrons and muons) have to be present and they enter the system as free gases. Therefore, charge neutrality and $\beta$-equilibrium conditions require that:
\begin{equation}
\mu_p = \mu_n -\mu_e, \quad \mu_e =\mu_\mu, 
\quad 
\rho_p = \rho_e + \rho_\mu.
\label{conditionH}
\end{equation}
The energy density and pressure for the leptons are given by:
\begin{equation}
\varepsilon_l = \frac{3}{\pi^{2}}  \int^{K_{F_{l}}}_{0}k^2 (m_{l}^2 +k^2)^{1/2}dk,
\label{leptonener}
\end{equation}
and 
\begin{equation}
p = \frac{1}{3 \pi^{2}} \int^{K_{F_{l}}}_{0}\frac{k^4}{(m_{l}^2 +k^2)^{1/2}}dk,
\label{leptonpress}
\end{equation}
where $K_{F_{l}}$ is the Fermi momentum for leptons and the electron and muon mass values are 0.511 MeV and 105.66 MeV, respectively.

The electron and muon densities read:
\begin{equation}
\rho_l={K_{F_l}^3}/{3\pi^2}.
\end{equation}

There are many possible parameterizations of the QHD model. We chose the IU--FSU parameterization proposed in \cite{iufsu}. Besides the tests performed in \cite{Dutra2014,Dutra2015}, IU--FSU is also successful in explaining the recent constraint that comes from the GW170817 observation \cite{Lourenco2018}. 

For the outer crust of the neutron star, it is necessary to use a model that describes the nuclear matter for the low density region. For this, we use the full EoS BPS \cite{bps}. 

\subsection{The Quark--Meson Coupling Model}

Now, we present the second relativistic EoS used in the present work to model the nuclear matter in NS. For this we use the quark--meson coupling (QMC) model \cite{guichon}.

In the QMC model, nucleons in nuclear medium are considered as a system of non--overlapping MIT bags \cite{mitbag}, where quarks within one bag interact with quarks in another bag through the exchange of scalar $(\sigma)$ and vector $(\omega,\rho)$ mesons. The usual RMF approximation is used to treat the meson fields.

The quark field, $\psi_{q}$, inside the bag then satisfies the equation of motion:
\begin{equation}
\left[ i{\partial\mkern-7.5mu/} - (m_{q}^{0} - g_{\sigma}^{q}) - g_{\omega}^{q}\omega\gamma^{0} + \frac{1}{2}g_{\rho}^{q}\tau_{z}\rho_{03}\gamma^{0}\right] \psi_{qB}(x) = 0,  
\end{equation}
where $q$ are the quarks \textit{(q = u, d}) of mass $m_{q}^{0}$, $\tau_{z}$ is the spin projection, and $g_{\sigma }^{q}$, $g_{\omega}^{q}$ and $g_{\rho}^{q}$ denote the quark--meson coupling constants. The energy of the static bag describing a nucleon ($p$ or $n$) consisting of three quarks, in the ground state, is expressed as
\begin{equation}
    E^{bag}_{p,n} = \sum_{q}n_{q}\frac{\Omega_{q_{p,n}}}{R_{p,n}} -\frac{Z_{p,n}}{R_{p,n}} + \frac{4}{3}\pi R_{p,n}^3B_{p,n},
 \end{equation}
where $Z_{p,n}$ is a parameter containing information about zero--point motion of nucleon and $B_{p,n}$ is the bag constant of radius $R_{p,n}$. The effective mass of a nucleon is defined as $M^{*}_{p,n} = E^{bag}_{p,n}$.

The equilibrium condition for the bag is obtained by minimizing the effective mass $M^{*}_{p,n}$ with respect to the bag radius
\begin{equation}
    \frac{d M^{*}_{p,n}}{d R^{*}_{p,n}} = 0.
\end{equation}
In our calculations, we consider $Z_{p,n}= 4.0050668$ and $E^{1/4}_{p,n}= 210.85$ MeV. These values are obtained by fixing the bag radius $R_{p,n} = 0.6$ fm and considering the bare nucleon mass $M = 939$ MeV. For more details see ref. \cite{Grams:2015inf}.

After some analytical calculations that can be found in \cite{Grams:2015inf,Grams18} the following expressions for energy density and pressure are obtained, namely:
\begin{eqnarray}
   \mathcal{E} &=& \frac{1}{2}m^2_\sigma\sigma + \frac{1}{2}m^2_\omega\omega_0^2 +  \frac{1}{2}m^2_\rho\rho_{03}^2 + \sum_{p,n}  \frac{1}{\pi^{2}}\int_0^{{k_F}_{p,n}}k^{2}(k^{2}+M^{*2}_{p,n})^{1/2}dk,
   \label{energydensqmc}
\end{eqnarray}
and pressure,
\begin{eqnarray}
  P &=& - \frac{1}{2}m^2_\sigma\sigma + \frac{1}{2}m^2_\omega\omega_0^2 + \frac{1}{2}m^2_\rho\rho_{03}^2 + \sum_{p,n} \frac{1}{3\pi^{2}}\int_0^{{k_F}_{p,n}}\frac{k^{4}dk}{(k^{2}+M^{*2}_{p,n})^{1/2}}.
  \label{pressureqmc}
\end{eqnarray}
The mesonic fields $\omega_{0}$ e $\rho_{03}$ are determined through the following relations
\begin{eqnarray}
  \omega_{0} = \frac{g_{\omega}(\rho_p + \rho_n)}{m^2_\omega} \ , \ \ \rho_{03} = \frac{g_{\rho}(\rho_p - \rho_n)}{m^2_\rho}, 
\end{eqnarray}
where
\begin{equation}
    \rho_{p,n} = \sum_{p,n}\frac{\gamma}{6\pi^2}{k_F^3}_{p,n},
    \label{barionic_dens}
\end{equation}
is the baryonic density. 

Again, charge neutrality and chemical equilibrium conditions need to be implemented, which depend on the inclusion of leptons. The leptonic expressions for energy density, pressure and density are the same as given in the last subsection. 

Once again, we use the full BPS EoS to describe the star outer crust \cite{bps}.

\section{Results} \label{results}

In this work we obtained sequences of solutions for four rotation regimes: non--rotating stars which are represented by dashed curves in the plots, stars rotating at $300$ Hz represented by dash double--dot curves, stars rotating at $716$ Hz shown in the dash--dot curves and stars rotating at the Kepler limit shown in the solid line curves. The Keplerian limit is reached when the angular velocity of the star at the equator $\Omega(R_e,\theta=\pi/2)$ is the same as the angular velocity $\Omega_K$ of a free particle in circular orbit
\begin{equation}
\Omega_{K}=\left.\left(\omega+\frac{r\partial_{r}\omega}{2+r\partial_{r}\gamma-r\partial_{r}\rho}+\sqrt{\left(\frac{r\partial_{r}\omega}{2+r\partial_{r}\gamma-r\partial_{r}\rho}\right)^{2}+\frac{e^{2\rho}(\partial_{r}\gamma+\partial_{r}\rho)}{r(2+r\partial_{r}\gamma-r\partial_{r}\rho)}}\;\right)\right|_{r=R_{e},\theta=\frac{\pi}{2}}.
\end{equation}
If we increase the rotation of the star beyond $\Omega_K$ the star starts to lose mass at the equator, so this limit is also known as the mass--shedding limit. Besides, we obtained solutions for four values of the parameter $\lambda'$ of the $f(R,T)$ theory: $0$, which recovers the GR case and is represented by the black color curves, $-0.02$ shown in the blue curves, $-0.04$ shown in the green curves and $-0.06$ represented by the red curves. We do not used values of $\lambda'>0$ because this leads to negative values for $p^{eff}$ and $\varepsilon^{eff}$, which violates the energy conditions.

\begin{figure}[t]
     \centering
     \begin{subfigure}[b]{0.49\textwidth}
         \centering
         \includegraphics[width=\textwidth]{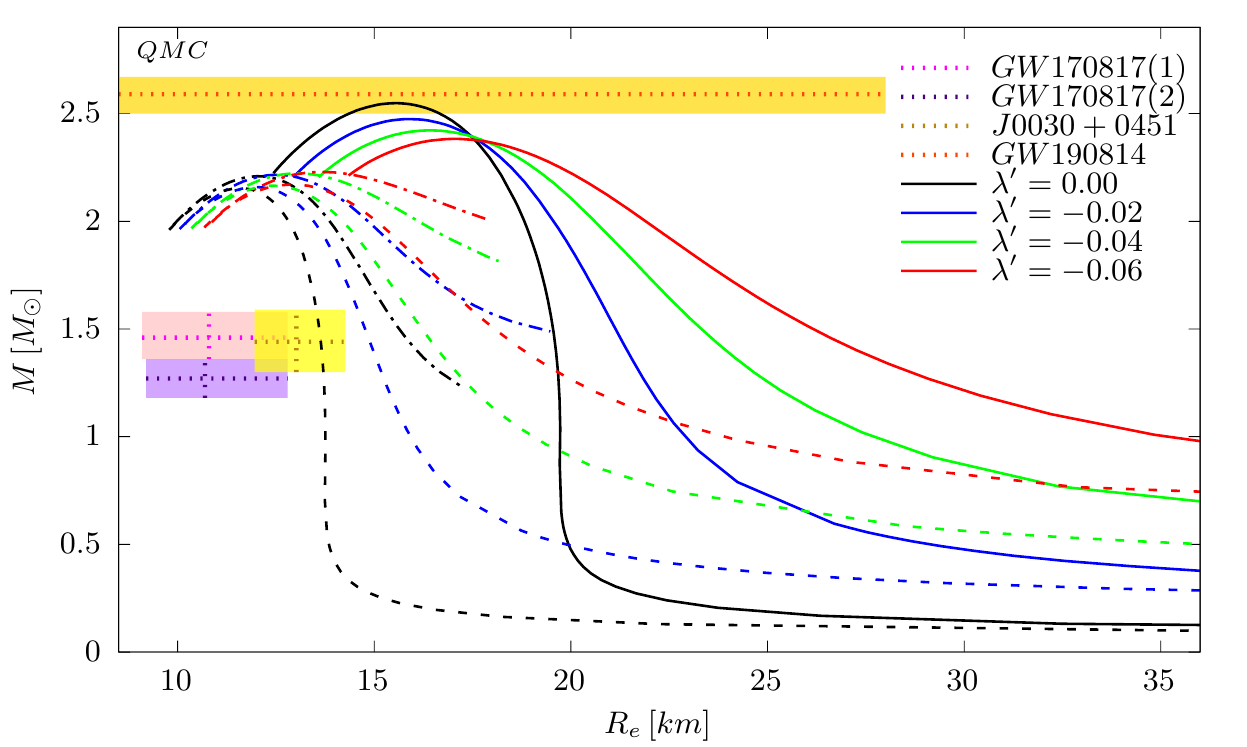}
     \end{subfigure}
     \begin{subfigure}[b]{0.49\textwidth}
         \centering
         \includegraphics[width=\textwidth]{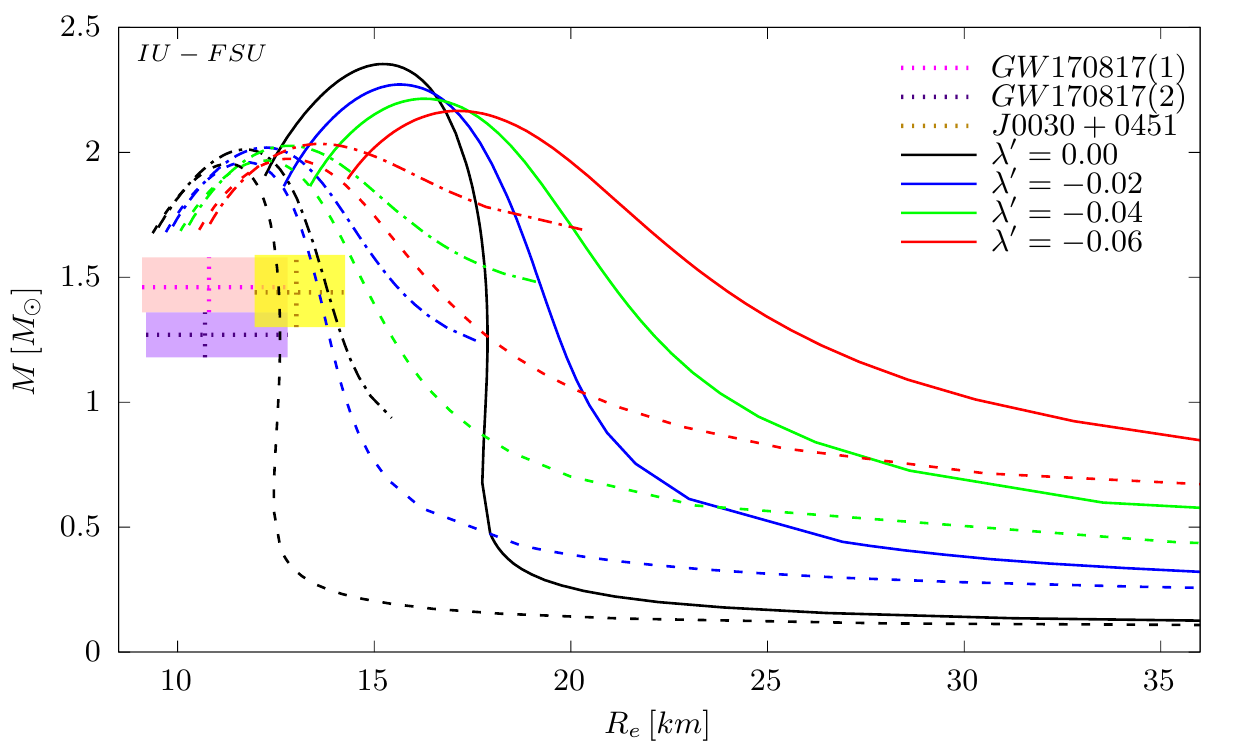}
     \end{subfigure}
        \caption{The mass--radius relation for sequences of non-rotating stars (dashed curves), stars rotating at $716 $ Hz (dash--dot curves) and for stars rotating at the mass--shedding limit (solid line curves). The curves for $\lambda'=0$ correspond to the GR case.}
        \label{fig1}
\end{figure}

Once we have obtained solutions for the field equations (\ref{rs21}), (\ref{rs16}), (\ref{rs17}) and (\ref{rs18}) together with the equation (\ref{heff}), which comes from the divergence of the energy momentum tensor, we are interested in analysing relevant physical quantities associated to the NS. The first analysis we do is for the mass--radius relation. For that, we need to calculate the gravitational mass of the star, also called total mass or tensor mass \cite{komatsu1989rapidly,cook1992spin,doneva2013rapidly}. This mass is obtained by the Komar integral \cite{komar1959covariant,wald2010general}, which is equal to the ADM mass \cite{arnowitt1959dynamical,arnowitt1960canonical,friedman2013rotating} in the case of stationary, asymptotically flat space--times. In the $f(R,T)$ theory such integral is given by
\begin{equation}
\begin{split}
M=&4\pi(1+2\lambda)\int_0^{\pi/2}\int_0^{R_{e}} e^{2\alpha+\frac{\gamma-\rho}{2}}\left[e^{\frac{\gamma+\rho}{2}}\left(\frac{\left(\varepsilon+p\right)\left(1+v^{2}\right)}{1-v^{2}}+2\frac{p(1+3\lambda)-\varepsilon\lambda}{1+2\lambda}\right)+\right. \\ &\left.2r  \sin\theta\;\omega e^{\frac{\gamma-\rho}{2}}\frac{\left(\varepsilon+p\right)v}{1-v^{2}}\right]r^{2}\sin\theta\; dr d\theta.
\end{split}
\end{equation}
The mass--radius relation is shown in Figure \ref{fig1} and, from these plots, we can conclude that the effect of the rotation is of increase the mass and the equatorial radius of the NS, in both GR and $f(R,T)$ gravity. We can also observe that the influence of the rotation on the mass is more intense for stars with the EoS IU--FSU than for those with QMC. For example, for the solutions with $\lambda'=-0.02$, when we go from non--rotating solutions to the ones rotating at $716$ Hz, there is an increase of $3.0\%$ in the maximum mass of the stars with EoS IU--FSU while for the stars with EoS QMC the increase is of $2.6\%$. The effect of the parameter $\lambda'$ on the mass--radius relation is that as we increase the absolute value of this parameter there is an increase in mass and equatorial radius for stars that are static and for those rotating at $716$ Hz. However, for the stars at the Kepler limit, the mass decreases as we increase the absolute value of $\lambda'$. This behavior is similar to that observed for Rastall gravity \cite{da2021rapidly} and for the dilatonic Einstein--Gauss--Bonnet theory \cite{kleihaus2016rapidly}. Another effect of $\lambda'$ that we can observe on the mass--radius relation is that as we increase the absolute value of this parameter, the curves gradually lose the characteristic of having several NS with almost the same radius in the astrophysically pertinent mass range. 

In Figure \ref{fig1} we show the masses of the two NS in the event GW170817 \cite{abbott2017gw170817,abbott2019gwtc}, respectively $M_1 =1.46_{-0.10}^{+0.12} M_{\odot}$ (pink--dotted horizontal line) and $M_2 =1.27_{-0.09}^{+0.09} M_{\odot}$ (purple--dotted horizontal line)\footnote{In Figure \ref{fig1}, the measurements for GW170817 and GW190814 are with $90 \%$ credible intervals and the ones for PSR J0030+0451 are with $68 \%$. The shaded bands in the plots indicate the intervals of each measurement.}. Using data from LIGO and Virgo \cite{abbott2018gw170817} it was estimated that the stars in this event have radii $R_1=10.8_{-1.7}^{+2.0}$ km (pink--dotted vertical line) and $R_2=10.7_{-1.5}^{+2.1}$ km (purple--dotted vertical line), respectively, with shaded regions to indicate the uncertainty in the measurements. We can observe that the static solutions in GR with EoS IU--FSU are in agreement with these data. However, for other sequences of solutions shown in Figure \ref{fig1} the radii for stars with $M_1$ and $M_2$ are larger than those estimated in \cite{abbott2018gw170817}. For example, the radii for NS with $M_1$ and $M_2$ which are rotating at $716$ Hz in $f(R,T)$ with $\lambda'=-0.02$ and EoS IU--FSU are $R_1 \approx 15.6$ km and $R_2 \approx 17.2$ km, respectively.

Figure \ref{fig1} also displays the estimates for the mass and the radius of the isolated $205.53 $ Hz millisecond pulsar PSR J0030+0451 made in \cite{miller2019psr}, which are $M_3 =1.44_{-0.14}^{+0.15} M_{\odot}$ (yellow--dotted horizontal line) and $R_3=13.02_{-1.06}^{+1.24}$ km (yellow--dotted vertical line). These estimates were made using data from NASA's Neutron Star Interior Composition Explorer (NICER) mission, installed on the International Space Station (ISS). NICER uses X--ray timing and spectroscopy instrument, to investigate soft X--ray emissions from hot spots on the surface of NS. Using the same data, independent estimates for PSR J0030+0451, which are compatible with those presented here, were also made in \cite{riley2019nicer}. We can observe that both EoS produce curves that have mass and radius values compatible with those estimated for PSR J0030+0451. However, for QMC EoS only the curve for static solutions in GR agrees with these data whereas for IU--FSU EoS we also have solutions in $f(R,T)$ with $\lambda'=-0.02$ the agree with $M_3$ and $R_3$.

Lastly, in the left panel of Figure \ref{fig1} we can see in orange--dotted horizontal line the mass of the compact object detected in the event GW190814 \cite{abbott2020gw190814} ($M_4 = 2.59_{-0.09}^{+0.08} M_{\odot}$). We can observe that stars rotating at the Kepler limit in GR with EoS QMC can achieve masses in the range of values of $M_4$, with a radius around $R_4 \approx 15.6$ km.

\begin{figure}[t]
     \centering
     \begin{subfigure}[b]{0.49\textwidth}
         \centering
         \includegraphics[width=\textwidth]{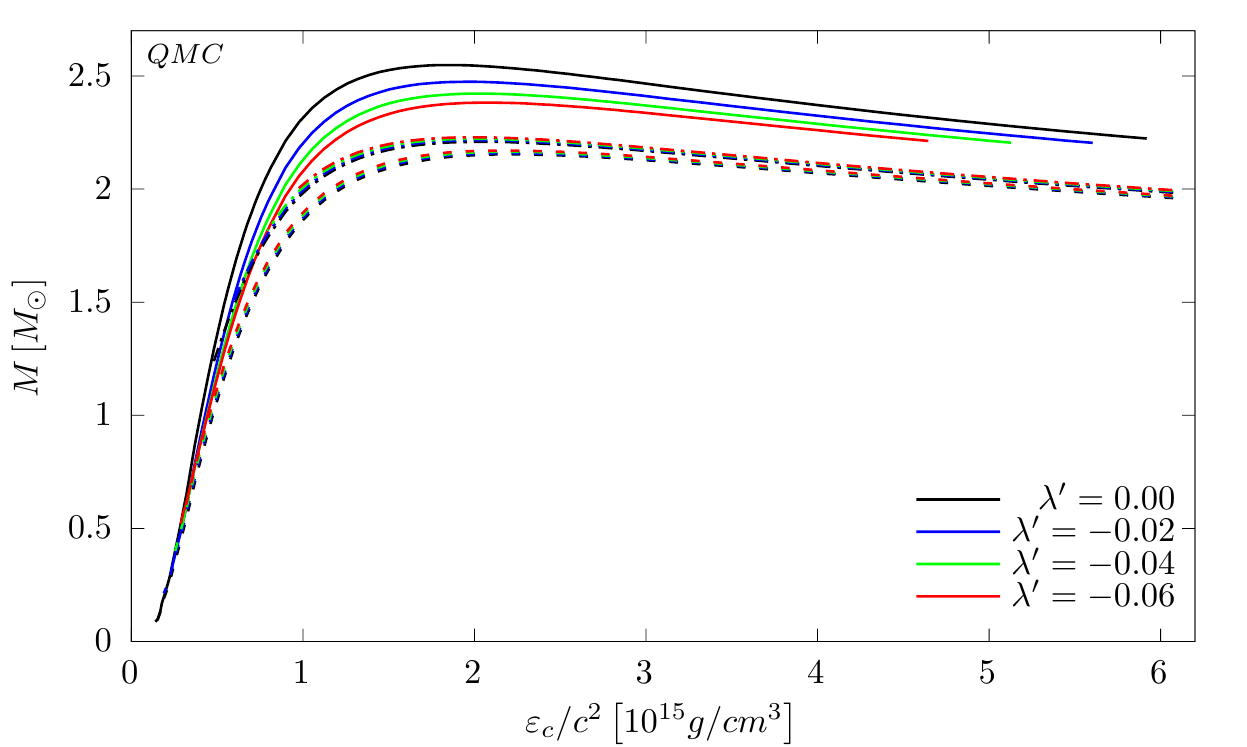}
     \end{subfigure}
     \begin{subfigure}[b]{0.49\textwidth}
         \centering
         \includegraphics[width=\textwidth]{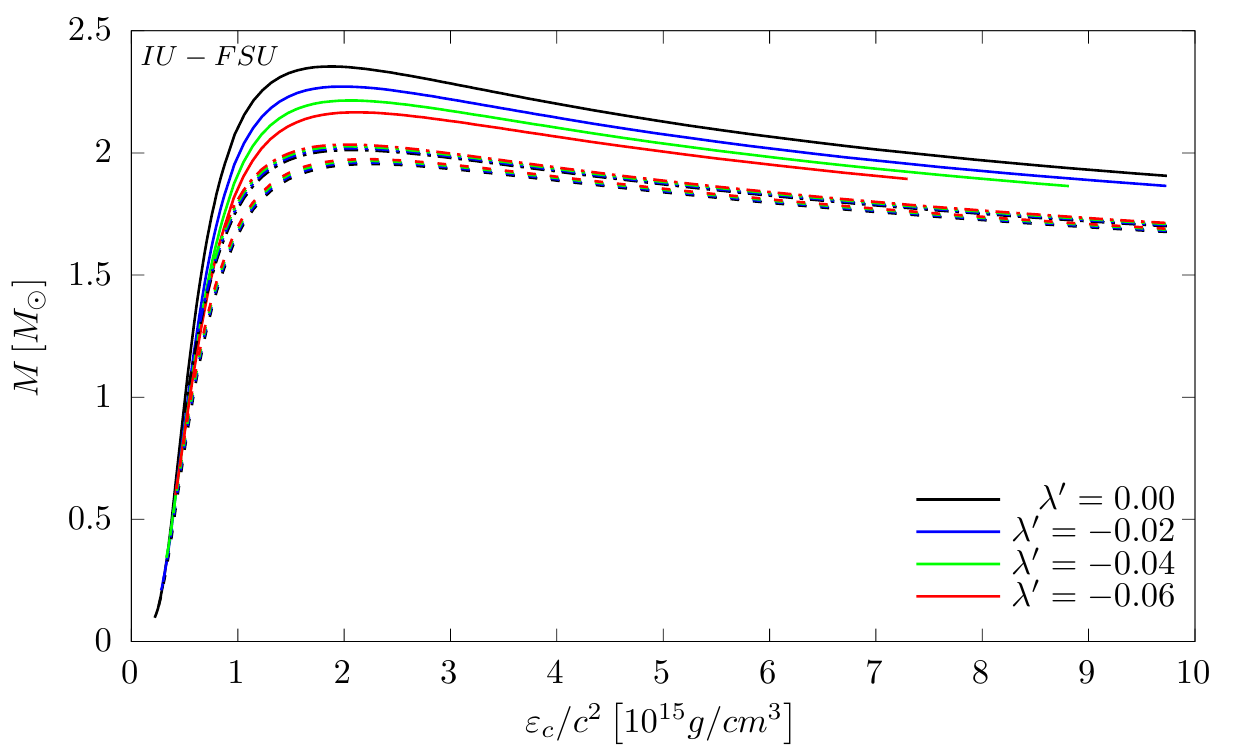}
     \end{subfigure}
        \caption{The mass as a function of the central energy density for sequences of non--rotating stars (dashed curves), stars rotating at $716$ Hz (dash--dot curves) and for stars rotating at the mass--shedding limit (solid line curves). The curves for $\lambda'=0$ correspond to the GR case.}
        \label{fig2}
\end{figure}

In Figure \ref{fig2} we show the relation between the mass and the central energy density. We can see that stars with higher angular velocity have higher masses for the same central energy density. As for the effect of the $f(R,T)$ gravity we can observe that as we increase the absolute value of $\lambda'$ there is an increase in the mass for the same value of $\varepsilon_c / c^2$ for stars that have $\Omega=0$ Hz and $\Omega=716$ Hz. But, for the solutions in the mass--shedding limit the effect of the parameter is of decrease the mass. We can observe that, in the Kepler limit, the maximum value of the central energy density decreases as we increase the absolute value of $\lambda'$. And, comparing the two plots we can note that for the EoS IU--FSU we can reach higher maximum central energy density than for EoS QMC.

Next, we are interested in the relation between the Kepler rotation and the angular momentum of the NS. The angular momentum is giver by the Komar integral as follows:
\begin{equation}
J=4\pi(1+2\lambda)\int_0^{\pi/2}\int_0^{R_{e}} \frac{\left(\varepsilon+p\right)v}{1-v^{2}} e^{2\alpha+\gamma-\rho} r^{3}\sin^{2}\theta\; dr d\theta.
\end{equation}
\begin{figure}[t]
     \centering
     \begin{subfigure}[b]{0.49\textwidth}
         \centering
         \includegraphics[width=\textwidth]{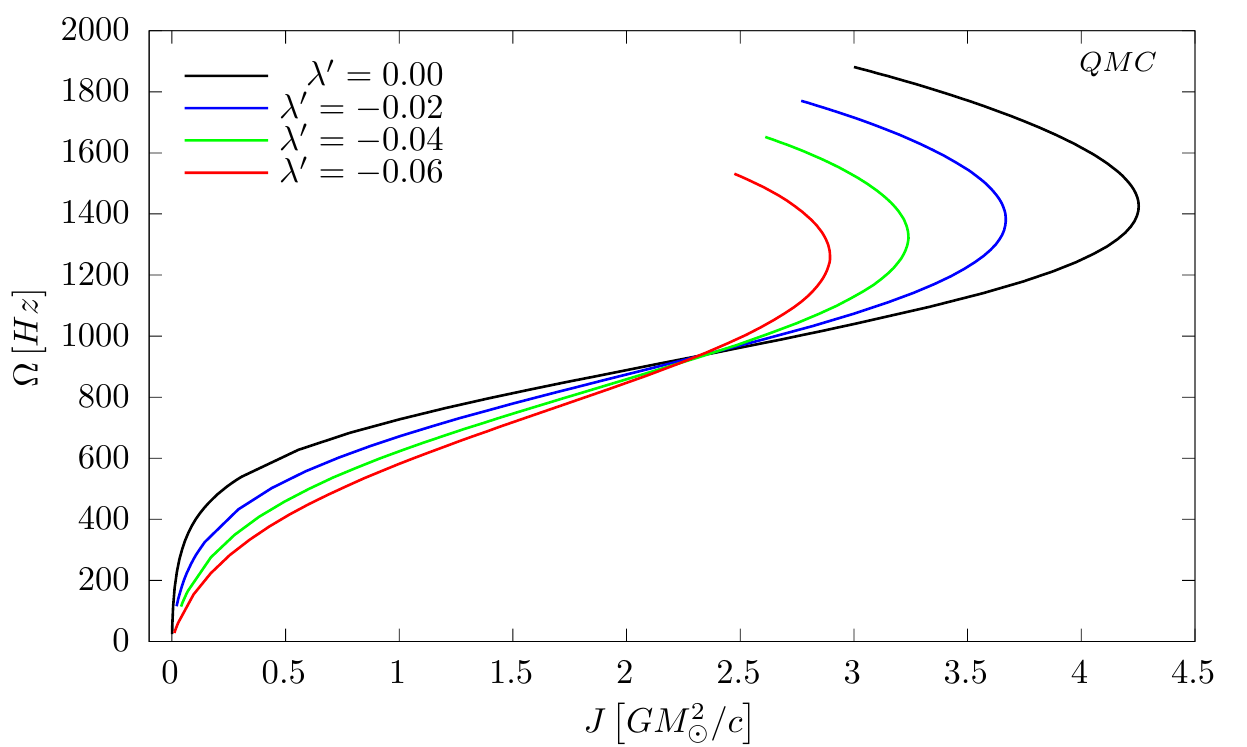}
     \end{subfigure}
     \begin{subfigure}[b]{0.49\textwidth}
         \centering
         \includegraphics[width=\textwidth]{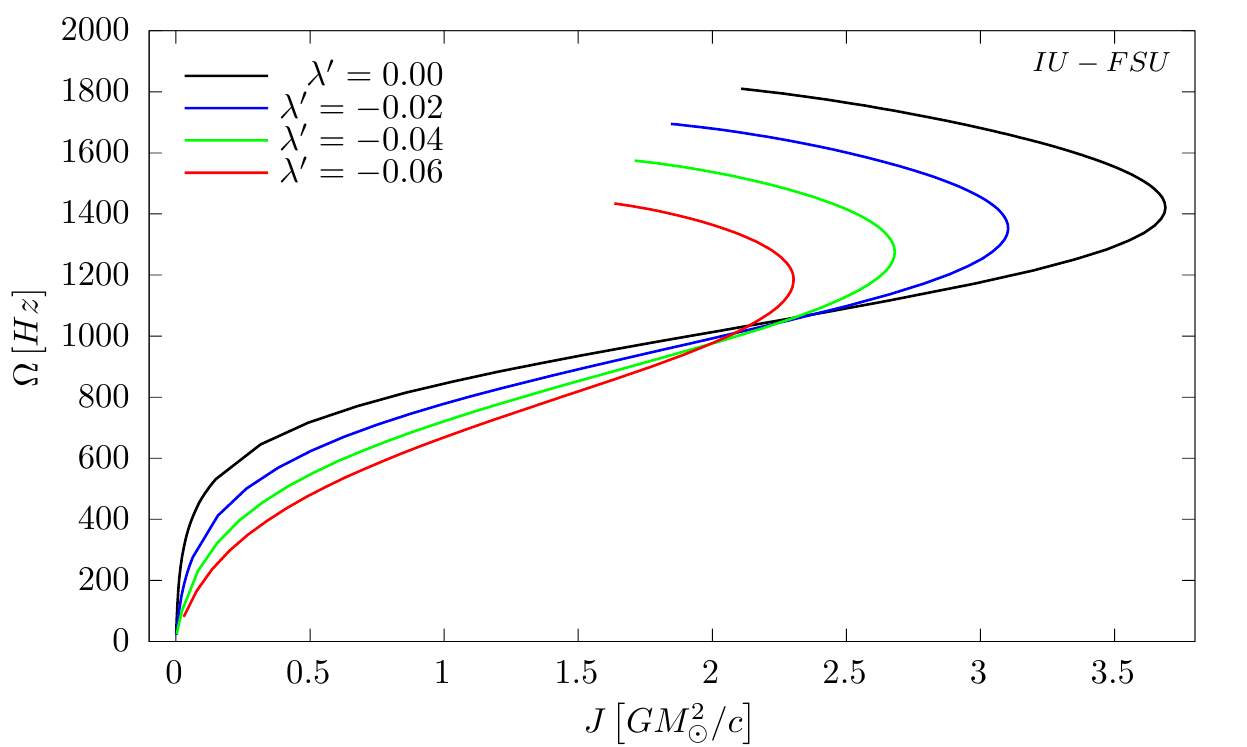}
     \end{subfigure}
        \caption{The angular velocity as a function of the angular momentum for sequences of stars rotating at the mass--shedding limit. The curves for $\lambda'=0$ correspond to the GR case.}
        \label{fig3}
\end{figure}
In Figure \ref{fig3} we plot the Kepler rotation versus the angular momentum. In this figure we can observe that the maximum angular momentum occurs to stars in GR and as we increase the value of the $f(R,T)$ parameter the maximum angular momentum decreases. The same occurs for the angular velocity, that is, as we increase the parameter $\lambda'$ the maximum velocity the star can rotate is decreased. We can also verify that NS with the EoS QMC reach higher values of $\Omega_K$ and $J$ than those with EoS IU--FSU.

Another quantity of interest is the moment of inertia, which can be found by calculating the ratio of the angular momentum $J$ to the angular velocity $\Omega$
\begin{equation}
    I=\frac{J}{\Omega}.
\end{equation}
\begin{figure}[t]
     \centering
     \begin{subfigure}[b]{0.49\textwidth}
         \centering
         \includegraphics[width=\textwidth]{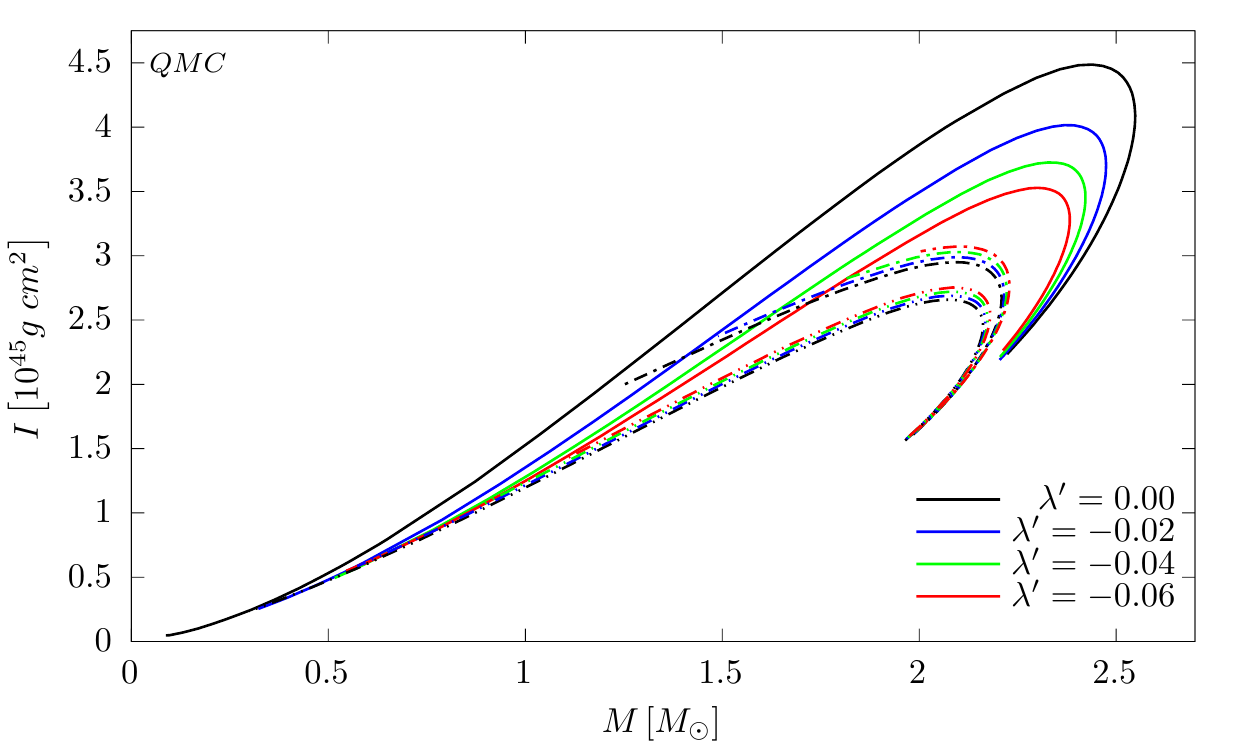}
     \end{subfigure}
     \begin{subfigure}[b]{0.49\textwidth}
         \centering
         \includegraphics[width=\textwidth]{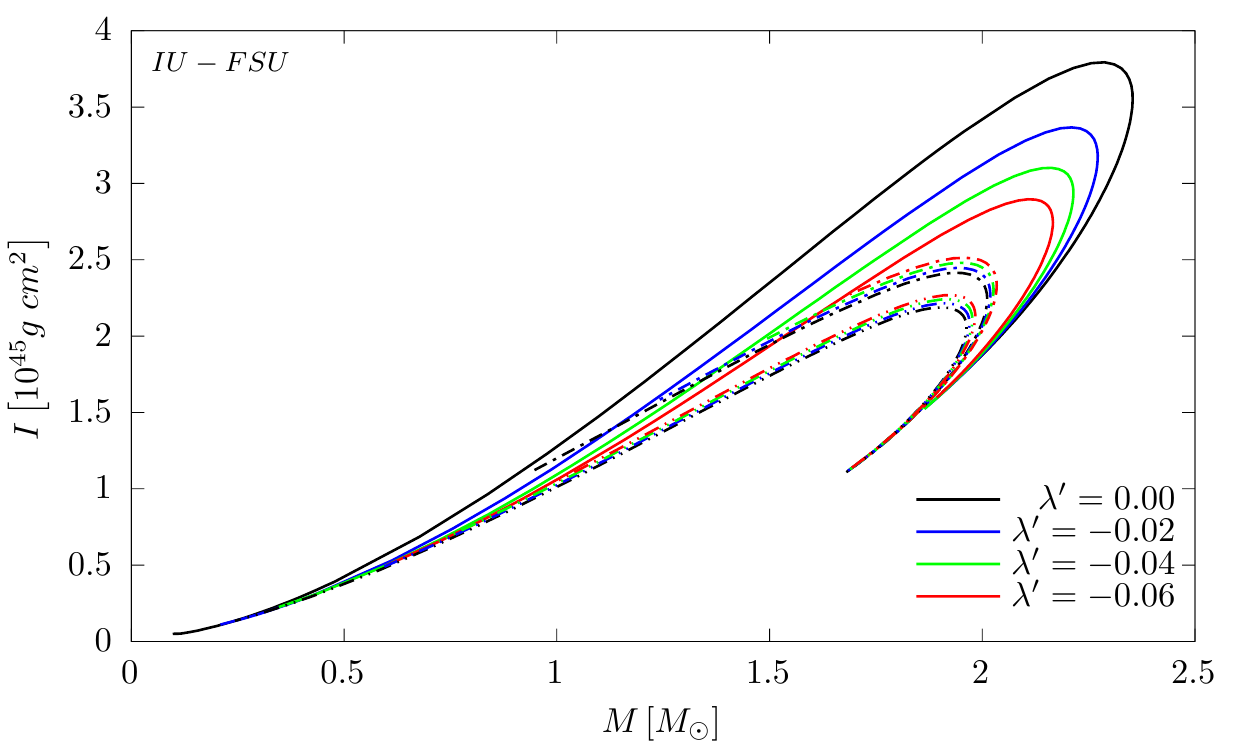}
     \end{subfigure}
        \caption{The Moment of inertia as a function of the mass for sequences of stars rotating at $300$ Hz (dash double--dot curves), stars rotating at $716 $ Hz (dash--dot curves) and for stars rotating at the mass--shedding limit (solid line curves). The curves for $\lambda'=0$ correspond to the GR case.}
        \label{fig4}
\end{figure}
This global parameter of the NS is very sensitive to the dense matter EoS, so that its determination has relevant implications to the constraining of the EoS models \cite{bejger2002moments,lattimer2005constraining,ozel2016masses}. And it is interesting to note that approximations for the moment of inertia can be constructed, especially for the more stiff EoS, also for non-rotating stars \cite{lattimer2000nuclear,tello2019anisotropic,singh2019minimally,singh2020static}. Figure \ref{fig4} shows the moment of inertia as a function of the mass. We can observe that for stars with $\Omega=300$ Hz and $\Omega=716$ Hz the moment of inertia increases as we increase the absolute value of $\lambda'$. However, for stars at the mass--shedding limit the maximum value for $I$ is attained for the GR case and, as we increase the absolute value of the parameter $\lambda'$ of the $f(R,T)$ gravity the values of the moment of inertia decrease. We can also see that NS with the EoS QMC achieve higher values for $I$ than those with EoS IU--FSU.

\begin{figure}[t]
     \centering
     \begin{subfigure}[b]{0.49\textwidth}
         \centering
         \includegraphics[width=\textwidth]{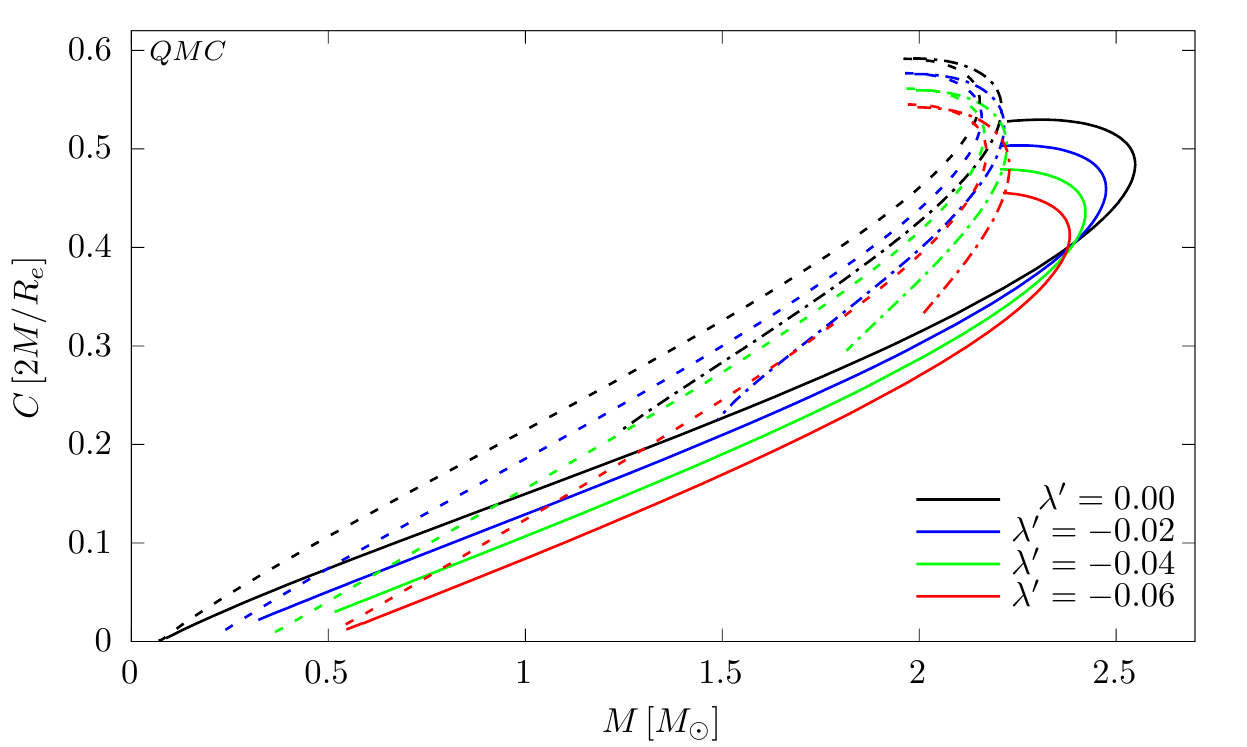}
     \end{subfigure}
     \begin{subfigure}[b]{0.49\textwidth}
         \centering
         \includegraphics[width=\textwidth]{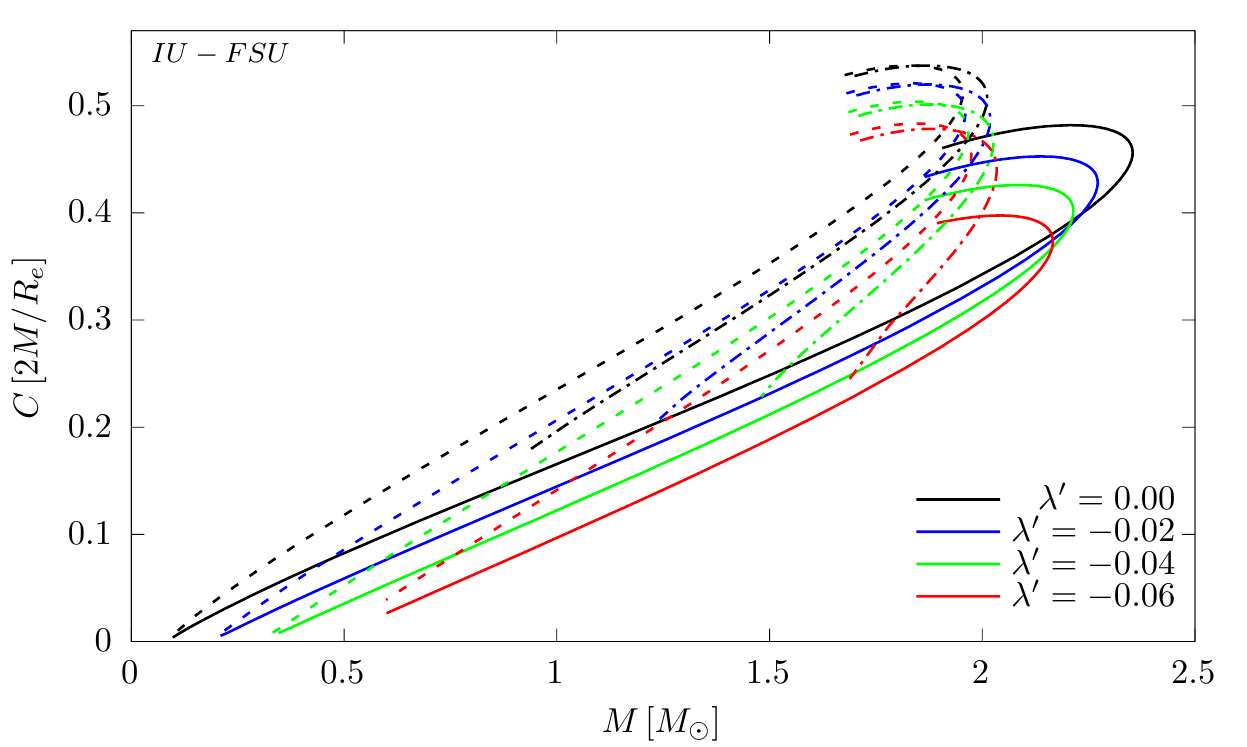}
     \end{subfigure}
        \caption{The compactness $C$ for non-rotating stars (dashed curves), stars rotating at $716 $ Hz (dash--dot curves) and for stars rotating at the mass--shedding limit (solid line curves). The curves for $\lambda'=0$ correspond to the GR case.}
        \label{fig5}
\end{figure}
The last physical quantity of interest we analyse is the compactness $C$ of the NS, it can be defined in terms of the mass $M$ and the equatorial radius $R_e$ as
\begin{equation}
    C=\frac{2M}{R_e}.
\end{equation}
This definition is the same one used in \cite{kleihaus2016rapidly}, and is normalized so that for black holes with mass $M$ and horizon radius $R_e$, we have $C=1$. In Figure \ref{fig5} we plot the compactness versus the mass for three different rotation regimes. We can observe that the effect of the $f(R,T)$ theory is of decrease the values of $C$. The plots show that the static NS are the ones with the highest values of compactness. And we can also note that stars with EoS IU--FSU achieve smaller values of compactness than the ones with EoS QMC.

\section{Conclusions} \label{conclusions}

In this work, we studied the influence of $f(R,T)$ theory on NS with realistic EoS in a fast rotation regime. We obtained results that indicate substantial modifications in the physical properties of NS in $f(R,T)$ gravity when compared to those in the context of GR. In particular, we have considered stars rotating at $0$ Hz, $300$ Hz, $716$ Hz and stars rotating at the Kepler limit. Regarding the $f(R,T)$ theory we considered four values of the parameter $\lambda'$: $0$, $-0.02$, $-0.04$ and $-0.06$. 

We obtained the mass--radius relation for sequences of stars and conclude that the presence of rotation increases the mass and the equatorial radius of the NS, in both GR and $f(R,T$) gravity. Besides, the effect of rotation is more intense in IU-–FSU model. Concerning the consequence of $f(R,T)$ gravity to the mass-radius relation, we have shown that the effect of increasing the values of the parameter $\lambda'$ is of increase the mass and equatorial radius for stars that are static and those rotating at $716$ Hz. But, for stars at the Kepler limit the mass decreases as we increase the value of the parameter $\lambda'$, similarly to what was observed in \cite{da2021rapidly} and \cite{kleihaus2016rapidly}. Using the mass and radius estimates for GW170817 and PSR J0030+0451 as constraints, we conclude that only values of $|\lambda'| \leq 0.02$ produce mass and radius values compatible with the experimental data considering the EoS used here. This bound for the parameter of $f(R,T)$ gravity is in agreement with \cite{lobato2020neutron}. As a complementary analysis of the mass--radius relation, we studied the relation between the mass and the central energy density. We were able to obtain solutions with higher values for the central energy density in the IU--FSU model than in the QMC model. Furthermore, in the mass shedding limit the maximum value for the central energy density decreases with the increasing of the parameter of the $f(R,T)$ gravity. 

We also examined the effect of the $f(R,T)$ theory in the relation between the Keplerian angular velocity and the angular momentum and concluded that this theory produces stars with smaller Kepler limit and smaller angular momentum than GR. Another physical quantity of great interest is the moment of inertia, which was investigated as a function of the mass in Figure \ref{fig4}. We found that for stars rotating at $300$ Hz and $716$ Hz the moment of inertia increases with the increasing of the parameter $\lambda'$, but for stars at the mass--shedding limit the opposite occurs. We also saw that the QMC model produces stars with higher values for the moment of inertia than the IU--FSU model.

Lastly, we studied the consequence of the $f(R,T)$ gravity to the compactness of the NS. We concluded that the effect of increasing the value of the parameter of this theory is of decrease the compactness of the stars both in the static and in the rotating case. 

In general, we could observe that our results for rapidly rotating stars in $f(R,T)$ gravity have some similarities with the ones obtained for Rastall's gravity \cite{da2021rapidly} and for dilatonic Einstein--Gauss--Bonnet theory \cite{kleihaus2016rapidly}.

\section{Acknowledgements}

We are very grateful to D.P. Menezes for providing the EoS tables used in this work. C.E.M. and T.O.F.C. are supported by Coordenação de Aperfeiçoamento de Pessoal de Nível Superior (CAPES) scholarship, L.C.N.S. would like to thank Conselho Nacional de Desenvolvimento Científico e Tecnológico (CNPq) for partial financial support through the research Project No. 164762/2020-5 and F.M.S. would like to thank CNPq for financial support through the research Project No. 165604/2020-4. J.C.F. thanks CNPq and FAPES for their financial support.

\bibliographystyle{ieeetr}
\bibliography{ref.bib}

\end{document}